\documentclass[english,british]{jfm}
\usepackage[T1]{fontenc}
\usepackage[latin9]{inputenc}
\usepackage{verbatim}
\usepackage{soul}
\usepackage{xcolor}
\usepackage{graphicx}
\usepackage{stmaryrd}
\usepackage{amssymb}
\usepackage{amsmath}
\usepackage{esint}
\usepackage{units}
\usepackage[authoryear]{natbib}
\PassOptionsToPackage{normalem}{ulem}
\usepackage{ulem}
\usepackage{babel}

\makeatletter

\providecolor{lyxadded}{rgb}{0,0,1}
\providecolor{lyxdeleted}{rgb}{1,0,0}

\def\vec#1{\boldsymbol{#1}}

\makeatother

\begin{document}

\title[Fully conservative hydraulic jumps and solibores in two-layer Boussinesq fluids]
{Fully conservative hydraulic jumps and solibores in two-layer Boussinesq fluids}

\author[J. Priede]{J\={a}nis Priede$^{1,2}$}
\affiliation{$^1$Centre for Fluids and Complex Systems,\\
Coventry University, UK\\
$^2$Department of Physics, University of Latvia, Riga, LV-1004, Latvia}

\maketitle

\begin{abstract}
We consider a special type of hydraulic jumps (internal bores) which,
in the vertically bounded system of two immiscible fluids with slightly
different densities, conserve not only the mass and impulse but also
the circulation and energy. This is possible only at specific combinations
of the upstream and downstream states. Two such combinations are identified
with arbitrary upstream and downstream interface heights. The first
has a cross symmetry between the interface height and shear on both
sides of the jump. This symmetry, which is due to the invariance of
the two-layer shallow-water system with swapping the interface height
and shear, ensures the automatic conservation of the impulse and energy
as well as the continuity of characteristic velocities across the
jump. The speed at which such jumps propagate is uniquely defined
by the conservation of the mass and circulation. The other possibility
is a marginally stable shear flow which can have fully conservative
jumps with discontinuous characteristic velocities. Both types of
conservative jumps are shown to represent a long-wave approximation
to the so-called solibores which appear as smooth permanent-shape
solutions in a weakly non-hydrostatic model. A new analytical solution
for solibores is obtained and found to agree very well with the previous
DNS results for partial-depth lock release flow. The finding that
certain large-amplitude hydraulic jumps can be fully conservative,
while most are not such even in the inviscid approximation, points
toward the wave dispersion as a primary mechanism behind the lossy
nature of internal bores.
\end{abstract}

\section{Introduction}

Hydraulic jumps are steep variations in the height of the liquid surface
which can propagate at a nearly constant speed over relatively large
distances. Such step-like gravity waves are often referred to as bores.
A well-known example is that of tidal bores \citep{Simpson1999}.
Hydraulic jumps can form also in stably stratified fluids where they
are known as internal bores \citep{Baines1995}. The latter occur
in various geophysical flows, such as coastal oceans \citep{Scotti2004}
and the inversion layers in the atmosphere \citep{Christie1992}.

One of the key characteristics of hydraulic jumps is their speed of
propagation. Determination of this speed theoretically is difficult
because the flow associated with bores tends to be very complex and
often turbulent. Notwithstanding this complexity, the speed at which
a bore propagates can be determined approximately by considering the
conservation of relevant quantities. In single fluid layers, the speed
of propagation is determined by the conservation of the mass and momentum
fluxes across the front of the bore. Application of the respective
conservation laws integrally to a box enclosing the jump constitutes
the basis of the hydraulic approximation, also known as the control
volume method \citep{Rayleigh1914}.

The same front speed is defined also by the Rankine-Hugoniot jump
conditions for the hydrostatic shallow-water (SW) equations governing
the conservation of mass and momentum in single fluid layers \citep{Whitham1974}.
The momentum, however, is not the only dynamical quantity that can
be conserved. In fact, the hydrostatic SW equations admit an infinite
number of locally conserved quantities \citep[p. 459]{Whitham1974}.
However, only one such quantity can in general be conserved besides
the mass. For example, the jumps conserving momentum do not in general
conserve energy. The preference for momentum over energy as a conserved
quantity is motivated by physical considerations. Namely, the loss
of energy can be attributed to the viscous dissipation in strongly
turbulent bores. Turbulence can also enhance vorticity by stretching
and tilting vortices so increasing the total amount of enstrophy \citep[Sec. 5.2, p. 270]{Batchelor1967}.
This can affect the conservation of the circulation flux across the
jump. However, there is no analogous physical mechanism by which turbulence
or viscosity could affect the flux of momentum across the jump.

The same considerations apply also to two-layer systems. However,
there are a few significant differences which concern two-layer systems
bounded by a rigid lid. The principal difference is a longitudinal
pressure gradient. It appears in this case due to the fixed total
height and ensures that the volumetric flux in one layer is equal
but opposite to that in the other layer. This makes the basic SW equations
for separate layers non-local. As a result, the longitudinal momentum
is not in general conserved in such systems \citep{Camassa2012}.
But it does not mean that the two-layer hydrostatic SW equations are
inherently non-local as it is often thought \citep{Fyhn2019}. These
equations can be represented in a number of locally conservative forms
which are mathematically equivalent as long as the waves are smooth
\citep{Priede2020}. Firstly, there is a circulation conservation
equation \citep{Sandstrom1993}. It yields the same speed of propagation
as the vorticity front condition obtained by \citet{Borden2013} using
the conventional control-volume approach. Secondly, there is also
a momentum-like quantity, called pseudo-momentum or impulse \citep{Benjamin1986},
which is conserved locally in two-layer systems bounded by a rigid
lid. However, in contrast to the single-layer case, the respective
two-layer SW conservation equation is not uniquely defined. It contains
a free parameter $\alpha$ which defines the relative contribution
of each layer to the pressure at the interface \citep{Priede2020}.
This parameter, which is expected to depend on the density ratio,
affects only the hydraulic jumps but not continuous waves. The Rankine-Hugoniot
jump conditions resulting from the mass and impulse conservation equations
with $\alpha=1$ and $\alpha=-1$ are mathematically equivalent to
the classical front conditions for internal bores obtained by \citet{Wood1984}
and \citet{Klemp1997}, respectively. The latter also includes the
front condition of \citet{Benjamin1968} for gravity currents. The
vorticity front condition of \citet{Borden2013} is in turn recovered
in the limit $|\alpha|\rightarrow\infty.$ With $\alpha=0,$which
is suggested by symmetry considerations for Boussinesq fluids, the
predicted jump propagation velocities agree well with the available
\foreignlanguage{english}{numerical and experimental results }\citep{Priede2020}.

Since the speed of propagation predicted by the SW model depends on
$\alpha,$ circulation like energy is not in general conserved by
the jumps that conserve the mass and impulse. This is because the
related conservation laws are mutually equivalent only for smooth
(differentiable) solutions \citep{Whitham1974}. There are, however,
a few exceptions that correspond to the jumps whose speed of propagation
does not depend on $\alpha.$ Such jumps, which conserve all four
quantities, are the focus of the present paper. We show that these
fully conservative jumps represent a long-wave approximation to the
solibores which are permanent-shape smooth solutions appearing in
the weakly non-hydrostatic approximation \citep{Esler2011}. We also
obtain an exact analytical solution for such solibores and find a
good agreement with the numerical results of \citet{Khodkar2017}
for the partial-height lock-exchange flow in a two-layer system with
a small density contrast. 

The paper is organized as follows. In section \ref{sec:model}, we
formulate the problem and present the basic equations including the
generalized SW momentum equation. Jump conditions are introduced and
analysed in section \ref{sec:Jump}. In section \ref{sec:analytic},
we show that the characteristics of fully conservative jumps are the
same as those of solibores and obtain a relatively simple analytical
solution for the latter. The paper is concluded with section \ref{sec:Sum}
which contains a summary and discussion of the main results including
a comparison with numerical results.

\section{\label{sec:model}Two-layer SW model}

\begin{figure}
\begin{centering}
\includegraphics[width=0.5\columnwidth]{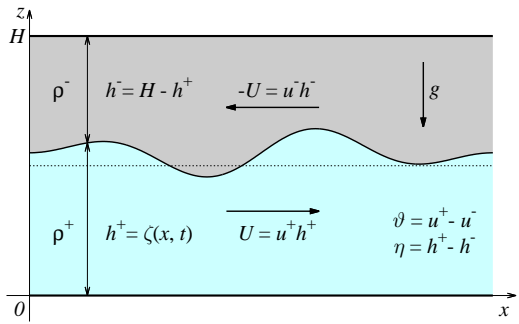}
\par\end{centering}
\caption{\label{fig:sktch1}Sketch of the problem showing a horizontal channel
of a constant height $H$ bounded by two parallel solid walls and
filled with two inviscid immiscible fluids with constant densities
$\rho^{+}$ and $\rho^{-},$ where $h^{+}=\zeta(x,t)$ and $h^{-}=H-h^{+}$are
the depths of the bottom and top layers, respectively.}
\end{figure}

Consider a horizontal channel of a constant height $H$ which is bounded
by two parallel solid walls and filled with two inviscid immiscible
fluids with constant densities $\rho^{+}$ and $\rho^{-}$ as shown
in Fig. \ref{fig:sktch1}. The fluids are subject to a downward gravity
force with the free fall acceleration $g.$ The interface separating
the fluids is located at the height $z=\zeta(x,t),$ which is equal
to the depth of the bottom layer $h^{+}$ and varies with the horizontal
position $x$ and the time $t.$  The velocity $\vec{u}^{\pm}$ and
the pressure $p^{\pm}$ in each layer are governed by the Euler equation
\begin{equation}
\partial_{t}\vec{u}+\vec{u}\cdot\vec{\nabla}\vec{u}=-\rho^{-1}\vec{\nabla}p+\vec{g}\label{eq:Euler}
\end{equation}
and the incompressibility constraint $\vec{\nabla}\cdot\vec{u}=0.$
Henceforth, for the sake of brevity, we drop $\pm$ indices wherever
analogous expressions apply to both layers. At the interface $z=\zeta(x,t)$,
we have the continuity of pressure, $[p]\equiv p^{+}-p^{-}=0,$ and
the kinematic condition 
\begin{equation}
w=\frac{d\zeta}{dt}=\zeta_{t}+u\zeta_{x},\label{eq:kin}
\end{equation}
where $\frac{d}{dt}$ denotes the material derivative, $u$ and $w$
are the $x$ and $z$ components of velocity, and the subscripts $t$
and $x$ stand for the corresponding partial derivatives.

To make the paper self-contained, below we shall present a brief derivation
of the basic shallow water equations. Firstly, integrating the incompressibility
constraint over the depth of each layer and using (\ref{eq:kin}),
we obtain
\begin{equation}
h_{t}+(h\bar{u})_{x}=0,\label{eq:hpm}
\end{equation}
where the overbar denotes the depth average. Secondly, doing the same
for the horizontal $(x)$ component of (\ref{eq:Euler}), we have
\begin{equation}
(h\bar{u})_{t}+(h\overline{u^{2}})_{x}=-\rho^{-1}h\overline{p}_{x}.\label{eq:ub}
\end{equation}
Pressure is obtained by integrating the vertical $(z)$ component
of (\ref{eq:Euler}) as follows
\begin{equation}
p(x,z,t)=\mathit{\Pi}(x,t)+\rho\int_{\zeta}^{z}(w_{t}+\vec{u}\cdot\vec{\nabla}w-g)dz,\label{eq:prs}
\end{equation}
where $\mathit{\Pi}(x,t)=\left.p^{\pm}(x,z,t)\right|_{z=\zeta}$ is
the distribution of pressure along the interface. Thirdly, averaging
the $x$-component of the gradient of the pressure (\ref{eq:prs})
over the depth of each layer, after a few rearrangements, we obtain
\begin{equation}
\overline{p}_{x}=\left(\mathit{\Pi}+\rho g\zeta+\rho\overline{(z-z_{0})(w_{t}+\vec{u}\cdot\vec{\nabla}w)}\right)_{x},\label{eq:pxb}
\end{equation}
which defines the RHS of (\ref{eq:ub}) with $z_{0}=0$ and $z_{0}=H$
for the bottom and top layers, respectively.

When the characteristic horizontal length scale $L$ is much larger
than the height $H:$ $H/L=\epsilon\ll1,$ the exact depth-averaged
equations obtained above can be simplified using the shallow-water
approximation. In this case, the incompressibility constraint implies
$w/u=O(\epsilon)$ and (\ref{eq:pxb}) correspondingly reduces to
\begin{equation}
\overline{p}_{x}=(\mathit{\Pi}+\rho g\zeta)_{x}+O(\epsilon^{2}),\label{eq:px0}
\end{equation}
where the leading-order term is purely hydrostatic and $O(\epsilon^{2})$
represents a small dynamical pressure correction due to the vertical
velocity $w.$ Besides that the flow in both layers is considered
to be irrotational: $\vec{\omega}=\vec{\nabla}\times\vec{u}=0.$ According
to the inviscid vorticity equation $\frac{d\vec{\omega}}{dt}=(\vec{\omega}\cdot\vec{\nabla})\vec{u},$
this property is preserved by (\ref{eq:Euler}). In the leading-order
approximation, the irrotationality constraint reduces to $\partial_{z}u^{(0)}=0$.
It means that the horizontal velocity can be written as 
\begin{equation}
u=\bar{u}+\tilde{u},\label{eq:ubt}
\end{equation}
where the deviation from the average $\bar{u}$ implied by (\ref{eq:px0})
is $\tilde{u}=O(\epsilon^{2}).$ Consequently, in the second term
of (\ref{eq:ub}), we can substitute $\overline{u^{2}}=\bar{u}^{2}+O(\epsilon^{4}).$
Finally, using (\ref{eq:hpm}) and ignoring the $O(\epsilon^{2})$
dynamical pressure correction, (\ref{eq:ub}) can be written as 
\begin{equation}
\rho\left(\bar{u}_{t}+\tfrac{1}{2}\text{\ensuremath{\bar{u}^{2}}}_{x}+g\zeta_{x}\right)=-\mathit{\Pi}_{x}.\label{eq:upm}
\end{equation}
This equation along with (\ref{eq:hpm}) constitute the basic set
of SW equations in the leading-order (hydrostatic) approximation.

The vertical velocity, which gives rise to a non-hydrostatic pressure
perturbation, follows from the incompressibility constraint and (\ref{eq:ubt})
as 
\begin{equation}
w(z)=-\int_{z_{0}}^{z}u_{x}dz=-(z-z_{0})\bar{u}_{x}+O(\epsilon^{2}),\label{eq:w}
\end{equation}
where $z_{0}$ is defined as in (\ref{eq:pxb}) to satisfy the impermeability
conditions $w(0)=w(H)=0.$ Substituting this into (\ref{eq:pxb}),
after a few transformations, we obtain the well-known result for the
first-order pressure correction \citep{Green1976,Liska1995,Choi1999}
\begin{equation}
h\overline{p}_{x}^{(1)}=-\tfrac{1}{3}\rho\left(h^{3}(D_{t}\bar{u}_{x}-\text{\ensuremath{\bar{u}_{x}}}^{2}\right)_{x}+O(\epsilon^{4})=\tfrac{1}{3}\rho\left(h^{2}D_{t}^{2}h\right)_{x}+O(\epsilon^{4}),\label{eq:mcc}
\end{equation}
where $D_{t}\equiv\partial_{t}+\bar{u}\partial_{x}$ and $\bar{u}_{x}=-h^{-1}D_{t}h.$
The latter relation follows from (\ref{eq:hpm}) and ensures that
(\ref{eq:kin}) is satisfied by (\ref{eq:w}) up to $O(\epsilon^{2}).$

The system of four SW equations (\ref{eq:upm}) and (\ref{eq:hpm})
contains five unknowns: $u^{\pm}$, $h^{\pm}$ and $\mathit{\Pi},$
and is completed by adding the fixed height constraint $\{h\}\equiv h^{+}+h^{-}=H.$
Henceforth, we simplify the notation by omitting the bar over $u$
and using the curly brackets to denote the sum of the enclosed quantities.

Two more unknowns are eliminated as follows. Firstly, adding the mass
conservation equations for each layer together and using $\{h\}_{t}\equiv0,$
we obtain $\{uh\}=\Phi(t),$ which is the total flow rate. The channel
is assumed to be laterally closed, which means $\Phi\equiv0,$ and,
thus, $u^{-}h^{-}=-u^{+}h^{+}.$ Secondly, the pressure gradient $\mathit{\Pi}_{x}$
can be eliminated by subtracting the two equations (\ref{eq:upm})
one from another. This leaves only two unknowns, $U\equiv u^{+}h^{+}$
and $h=h^{+},$ and two equations, which can be written in a locally
conservative form as 
\begin{align}
\left(\left\{ \rho/h\right\} U\right)_{t}+\left(\tfrac{1}{2}\left[\rho/h^{2}\right]U^{2}+g\left[\rho\right]h\right)_{x} & =-\left[\overline{p}_{x}^{(1)}\right],\label{eq:u}\\
h_{t}+U_{x} & =0.\label{eq:h}
\end{align}
where the square brackets denote the difference of the enclosed quantities
between the bottom and top layers: $\left[f\right]\equiv f^{+}-f^{-}.$
The pressure correction on the RHS of (\ref{eq:u}), which is defined
by (\ref{eq:mcc}), can be cast in the following locally conservative
form 
\begin{equation}
\overline{p}_{x}^{(1)}=\frac{\rho}{3}\left(\left(hD_{t}^{2}h+\tfrac{1}{2}(D_{t}h)^{2}-h_{t}D_{t}h\right)_{x}+\left(h_{x}D_{t}h\right)_{t}\right).\label{eq:p1x}
\end{equation}
In the hydrostatic approximation, which will be considered first,
this dynamical pressure correction is irrelevant.

In the following, the density difference is assumed to be small. According
to the Boussinesq approximation, this difference is important only
for the gravity of fluids, which drives the flow. For the inertia,
this difference is ignored, which simplifies the problem significantly.
A further simplification is achieved by using the total height $H$
and the characteristic gravity wave speed $C=\sqrt{2Hg[\rho]/\{\rho\}}$
as the length and velocity scales, respectively, and $H/C$ as the
time scale.

In the hydrostatic Boussinesq approximation, (\ref{eq:u}) and (\ref{eq:h})
take a remarkably symmetric form \citep{Milewski2015}:
\begin{align}
\vartheta_{t}+\tfrac{1}{2}(\eta(1-\vartheta^{2}))_{x} & =0,\label{eq:crc}\\
\eta_{t}+\tfrac{1}{2}(\vartheta(1-\eta^{2}))_{x} & =0,\label{eq:vlm}
\end{align}
where $\eta=[h]$ and $\vartheta=[u]$ are the dimensionless depth
and velocity differentials between the top and bottom layers. Subsequently,
the former is referred to as the interface height and the latter as
the shear velocity. Note that this basic system of two-layer SW equations
is invariant to swapping $\eta$ and $\vartheta.$ The same symmetry
holds also for the momentum and energy equations 
\begin{align}
(\eta\vartheta)_{t}+\tfrac{1}{4}(\eta^{2}+\vartheta^{2}-3\eta^{2}\vartheta^{2})_{x} & =0,\label{eq:mnt}\\
(\eta^{2}+\vartheta^{2}-\eta^{2}\vartheta^{2})_{t}+(\eta\vartheta(1-\eta^{2})(1-\vartheta^{2}))_{x} & =0,\label{eq:nrg}
\end{align}
which are obtained by multiplying (\ref{eq:crc}) with $\eta$ and
$\eta^{2}\vartheta,$ respectively, and then using (\ref{eq:vlm})
to convert the resulting expression into locally conservative form.
It is important to note that this $\eta\leftrightarrow\vartheta$
symmetry is limited to the Boussinesq approximation. Also, note that
the conserved quantity $\eta\vartheta$ in (\ref{eq:mnt}) represents
a pseudo-momentum (impulse) \citep{Priede2020}. An infinite sequence
of further conservation laws can be constructed in a similar way \citep{Milewski2015}.
The generalized momentum equation can formally be obtained by multiplying
(\ref{eq:crc}) with an arbitrary constant $\alpha$ and adding to
(\ref{eq:mnt}):

\begin{equation}
((\eta+\alpha)\vartheta)_{t}+\tfrac{1}{4}(\eta^{2}+\vartheta^{2}-3\eta^{2}\vartheta^{2}+2\alpha\eta(1-\vartheta^{2}))_{x}=0.\label{eq:gen}
\end{equation}
For a more detailed derivation of equations (\ref{eq:mnt})--(\ref{eq:gen}),
we refer to \citet{Priede2020}. \foreignlanguage{english}{It has
to be stressed that equations (}\ref{eq:crc}\foreignlanguage{english}{)
and }(\ref{eq:mnt}\foreignlanguage{english}{)--(}\ref{eq:gen}\foreignlanguage{english}{)
are mutually equivalent and can be transformed one into another using
(}\ref{eq:vlm}\foreignlanguage{english}{) only if $\vartheta$ and
$\eta$ are differentiable. This, however, is not the case across
hydraulic jumps which will be considered in the next section. As the
problem is governed by two equations, only two corresponding jump
conditions can in general be satisfied.}

The constant $\alpha$ in (\ref{eq:gen}), which defines the relative
contribution of each layer to the pressure gradient along the interface,
is supposed to depend only on the ratio of densities. For nearly equal
densities, which is the case covered by the Boussinesq approximation,
$\alpha\approx0$ can be expected. This corresponds to both layers
affecting the pressure at the interface with equal weight coefficients.
Note that with $\alpha=0$ (\ref{eq:gen}) reduces to the basic momentum
equation (\ref{eq:mnt}) so restoring the $\eta\leftrightarrow\vartheta$
symmetry of the Boussinesq approximation. This symmetry is recovered
also in the limit |$\alpha|\rightarrow\infty,$ in which (\ref{eq:gen})
reduces to the circulation conservation equation (\ref{eq:crc}).

Two-layer SW equations for Boussinesq fluids can also be written in
the canonical form 
\begin{equation}
R_{t}^{\pm}-\lambda^{\pm}R_{x}^{\pm}=0,\label{eq:canon}
\end{equation}
where 
\begin{equation}
R^{\pm}=-\eta\vartheta\pm\sqrt{(1-\eta^{2})(1-\vartheta^{2})}\label{eq:Ri}
\end{equation}
 are the Riemann invariants and 
\begin{equation}
\lambda^{\pm}=\tfrac{3}{4}R^{\pm}+\tfrac{1}{4}R^{\mp}=-\eta\vartheta\pm\tfrac{1}{2}\sqrt{(1-\eta^{2})(1-\vartheta^{2})}\label{eq:Cpm}
\end{equation}
are the associated characteristic velocities \foreignlanguage{english}{\citep{Long1956,Cavanie1969,Ovsyannikov1979,Sandstrom1993,Baines1995,Chumakova2009}}.

For the interface confined between the top and bottom boundaries,
which corresponds to $\eta^{2}\le1,$ the characteristic velocities
(\ref{eq:Cpm}) are real and, thus, the equations are of hyperbolic
type if $\vartheta^{2}\le1.$ This hyperbolicity constraint on the
shear velocity ensures the absence of the long-wave Kelvin-Helmholtz
instability which would otherwise disrupt the interface \citep{Milewski2004}.
It has to be noted that this instability is different from the usual
short-wave Kelvin-Helmholtz which does not appear in the hydrostatic
SW approximation \citep{Esler2011}.

\section{\label{sec:Jump}Hydraulic jumps}

\begin{figure}
\begin{centering}
\includegraphics[width=0.5\columnwidth]{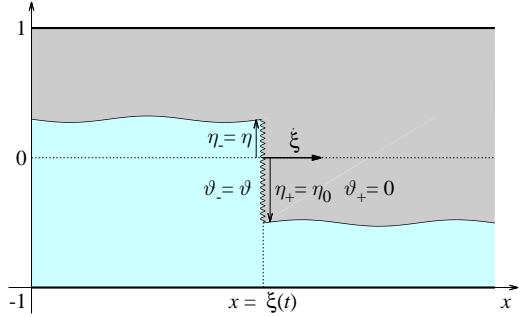}
\par\end{centering}
\caption{\label{fig:sktch2}A jump with the upstream interface height $\eta_{-}=\eta$
and the shear velocity $\vartheta_{-}=\vartheta$ propagating at the
speed $\dot{\xi}$ into a still fluid ahead $(\vartheta_{+}=0)$ with
the interface located at the height $\eta_{+}=\eta_{0}.$}
\end{figure}
Consider a discontinuity in $\eta$ and $\vartheta$ at the point
$x=\xi(t)$ across which the respective variables jump by $\left\llbracket \eta\right\rrbracket \equiv\eta_{+}-\eta_{-}$
and $\left\llbracket \vartheta\right\rrbracket \equiv\vartheta_{+}-\vartheta_{-}.$
Here the plus and minus subscripts denote the corresponding quantities
at the front and behind the jump. The double-square brackets stand
for the differential of the enclosed quantity across the jump. Integrating
(\ref{eq:vlm}) and (\ref{eq:gen}) across the jump, which is equivalent
to substituting the spatial derivative $f_{x}$ with $\left\llbracket f\right\rrbracket $
and the time derivative $f_{t}$ with $-\dot{\xi}\left\llbracket f\right\rrbracket $
\citep{Whitham1974}, the jump propagation velocity can be expressed,
respectively, as \citep{Priede2020}
\begin{eqnarray}
\dot{\xi} & = & \frac{1}{2}\frac{\left\llbracket \vartheta(1-\eta^{2})\right\rrbracket }{\left\llbracket \eta\right\rrbracket },\label{eq:jmp1}\\
\dot{\xi} & = & \frac{1}{4}\frac{\left\llbracket \eta^{2}+\vartheta^{2}-3\eta^{2}\vartheta^{2}+2\alpha(1-\vartheta^{2})\right\rrbracket }{\left\llbracket (\eta+\alpha)\vartheta\right\rrbracket }.\label{eq:jmp2}
\end{eqnarray}
For a jump to be feasible, it has to satisfy the hyperbolicity constraint
$\vartheta_{\pm}^{2}\le1$ as well as the energy constraint. The latter
follows from the integration of (\ref{eq:nrg}) across the jump and
defines the associated energy production rate \citep{Priede2020}
\begin{equation}
\dot{\varepsilon}=\left\llbracket \eta\vartheta(1-\eta^{2})(1-\vartheta^{2})\right\rrbracket -\dot{\xi}\left\llbracket \eta^{2}+\vartheta^{2}-\eta^{2}\vartheta^{2}\right\rrbracket \le0,\label{eq:deps}
\end{equation}
which cannot be positive as the energy can only be dissipated or dispersed
but not generated by the jump.

Applying (\ref{eq:jmp1}) and (\ref{eq:jmp2}) to a jump with the
upstream interface height $\eta_{-}=\eta$ propagating into a quiescent
fluid $(\vartheta_{+}=0)$ with the interface located at the height
$\eta_{+}=\eta_{0},$ as shown in Fig. \ref{fig:sktch2}, after a
few rearrangements, we obtain:

\begin{align}
\vartheta_{-}^{\pm} & =\pm\frac{(\eta_{0}-\eta)(\eta_{0}+\eta+2\alpha)^{1/2}}{((1-\eta^{2})(\eta_{0}-\eta)+2(\eta+\alpha)(1-\eta_{0}\eta))^{1/2}},\label{eq:v2}\\
\dot{\xi}^{\pm} & =-\vartheta_{-}^{\pm}\frac{1-\eta^{2}}{2(\eta_{0}-\eta)},\label{eq:xid}
\end{align}
where $\vartheta_{-}^{\pm}$ is the upstream shear velocity and the
plus and minus signs refer to the opposite directions of propagation,
i.e., $\dot{\xi}^{+}=-\dot{\xi}^{-},$ which are both permitted by
the mass and momentum balance conditions. Because the energy balance
(\ref{eq:deps}) changes sign with the direction of propagation, only
one direction of propagation is in general permitted for each jump.
This does not apply to the jumps which satisfy (\ref{eq:deps}) exactly.
Such energy-conserving jumps, which can in principle propagate downstream
as well as upstream, will be considered in the following.

A distinctive feature of these jumps is the invariance of their velocity
of propagation (\ref{eq:xid}) with $\alpha$ \citep{Priede2020}.
It implies that these jumps conserve also circulation. If this is
the case, the $\alpha$ terms in the generalized momentum equation
(\ref{eq:gen}) cancel out so making the associated jump condition
independent of $\alpha.$ For this to happen, the shear velocity (\ref{eq:v2})
at $\alpha=0$ has to be the same as that at $\alpha\rightarrow\infty.$
This requirement results in $\eta(\eta-\eta_{0})=0,$ which has two
solutions: $\eta=\eta_{0}$ and $\eta=0.$ The first one is irrelevant
as it corresponds to a uniform state with a constant interface height
$\eta_{0}.$ The second one describes a jump from $\eta_{0}$ to the
channel mid-height $\eta=0.$ The corresponding upstream shear velocity
and the speed of propagation following from (\ref{eq:v2}) and (\ref{eq:xid})
are $\vartheta_{-}^{\pm}=\pm\eta_{0}$ and $\dot{\xi}^{\pm}=\mp\frac{1}{2}.$

Owing to the symmetry of this jump $\vartheta_{\pm}=\eta_{\mp},$
the impulse and energy are conserved automatically, i.e., independently
of the velocity of propagation. Besides that the same symmetry makes
both characteristic velocities (\ref{eq:Cpm}) continuous across the
jump: $\left\llbracket \lambda^{\pm}\right\rrbracket =0.$ It means
that the positive, as well as negative characteristics, emanating
upstream from the jump are parallel to the corresponding characteristics
emanating downstream.  Likewise continuous are also the Riemann invariants
(\ref{eq:Ri}).

The above properties are shared by a broader class of fully conservative
jumps which can be sustained by a non-zero downstream shear $(\vartheta_{+}\not=0).$
The continuity condition $\left\llbracket \lambda^{\pm}\right\rrbracket =0$,
which after a few rearrangements can be rewritten as $\left\llbracket (\eta\pm\vartheta)^{2}\right\rrbracket =0,$
yields two pairs of possible solutions: 
\begin{align}
(\eta_{+},\vartheta_{+}) & =\pm(\eta_{-},\vartheta_{-}),\label{eq:jmp-s}\\
(\eta_{+},\vartheta_{+}) & =\pm(\vartheta_{-},\eta_{-}).\label{eq:jmp-a}
\end{align}
The first solution corresponding to the plus sign in (\ref{eq:jmp-s})
is continuous and so irrelevant. The other three are symmetric jumps
which automatically satisfy the energy balance condition (\ref{eq:deps})
as well as that of the impulse (\ref{eq:jmp2}) with $\alpha=0.$
 The second solution corresponding to the minus sign in (\ref{eq:jmp-s}),
which represents a centrally symmetric jump, conserves the mass and
circulation only if $\vartheta_{+}=\pm\eta_{+}$ and $\dot{\xi}=\pm\frac{1}{2}(1-\eta_{+}^{2}).$
This is just a particular case of a more general solution that follows
from the mass and circulation conservation laws for the second pair
of jumps (\ref{eq:jmp-a}). In this case, we obtain the speed of propagation
which can be written in terms of the upstream and downstream interface
heights as
\begin{equation}
\dot{\xi}^{\pm}=\pm\frac{1}{2}(1+\eta_{+}\eta_{-}).\label{eq:xipm}
\end{equation}
The corresponding shear velocities according to (\ref{eq:jmp-a})
are $(\vartheta_{+},\vartheta_{-})=\mp(\eta_{-},\eta_{+}).$ Note
that the previous two solutions with $\eta_{+}=0$ and $\eta_{+}=-\eta_{-}$
are particular cases of (\ref{eq:xipm}).

In marginally hyperbolic shear flows with $\vartheta_{+}=\vartheta_{-}=\mp1,$
which is a very specific case, fully conservative jumps with discontinuous
characteristic velocities are possible. Such jumps propagate at the
speed $\dot{\xi}^{\pm}=\pm\frac{1}{2}(\eta_{+}+\eta_{-})$ which ensures
the conservation of impulse and mass, whilst the energy and circulation
are conserved automatically.

In the next section, we show that these fully conservative hydraulic
jumps represent a long-wave approximation to the so-called solibores
which appear as permanent-shape solutions in the weakly non-hydrostatic
approximation described by (\ref{eq:mcc}) \citep{Esler2011}.

\section{\label{sec:analytic}Weakly non-hydrostatic analytical solution for
solibores}

Let us now turn to a weakly non-hydrostatic approximation, which is
defined by the dynamical pressure correction (\ref{eq:p1x}) in the
circulation conservation equation (\ref{eq:u}), and search for permanent-shape
waves travelling at the speed $c.$ Such waves are stationary in the
co-moving frame of reference where they vary depending only on $x'=x-ct.$
With the time derivative written as $\partial_{t}\equiv-c\partial_{x},$
the last two terms in (\ref{eq:p1x}) cancel out, and correspondingly
(\ref{eq:u}) and (\ref{eq:h}) take the form
\begin{eqnarray}
\left(\tfrac{1}{2}\eta(1-\vartheta^{2})-c\vartheta\right)_{x} & = & -\tfrac{1}{3}\left[hD_{t}^{2}h+\tfrac{1}{2}(D_{t}h)^{2}\right]_{x},\label{eq:crc-tw}\\
\left(\tfrac{1}{2}\vartheta(1-\eta^{2})-c\eta\right)_{x} & = & 0,\label{eq:vlm-tw}
\end{eqnarray}
where $h$ stands for $h^{\pm}=\frac{1}{2}(1\pm\eta)$ and $D_{t}$
for $D_{t}^{\pm}\equiv(u^{\pm}-c)\partial_{x}$ with the plus and
minus indices referring to the top and bottom layers. Now (\ref{eq:crc-tw})
and (\ref{eq:vlm-tw}) can be integrated once to obtain 
\begin{eqnarray}
\tfrac{1}{2}\eta(1-\vartheta^{2})-c(\vartheta-\vartheta_{0}) & = & -\tfrac{1}{3}\left[hD_{t}^{2}h+\tfrac{1}{2}(D_{t}h)^{2}\right],\label{eq:crc-sol}\\
\tfrac{1}{2}\vartheta(1-\eta^{2})-c(\eta-\eta_{0}) & = & 0,\label{eq:vlm-sol}
\end{eqnarray}
where $\vartheta_{0}$ and $\eta_{0}$ are the constants of integration.
Note that if the fluid far up- or downstream is at rest $(\vartheta=0),$
then according to (\ref{eq:vlm-sol}) $\eta_{0}$ is equal to the
respective interface height; $c\vartheta_{0}=A$ is the flux of circulation
which is one of the conserved quantities.

Using the identity $hD_{t}^{\pm}\equiv-h_{0}^{\pm}c\partial_{x},$
which follows from $D_{t}^{\pm}\equiv(u^{\pm}-c)\partial_{x}$ and
(\ref{eq:vlm-sol}) when written in terms of the original variables
as $h^{\pm}(u^{\pm}-c)=-h_{0}^{\pm}c$, the RHS of (\ref{eq:crc-sol})
can be transformed as follows 
\[
-\frac{c^{2}}{3}\left[h_{0}^{2}\left(\left(h_{x}/h\right)_{x}+\tfrac{1}{2}\left(h_{x}/h\right)^{2}\right)\right]=-\frac{c^{2}}{3}\left\{ (h_{0}h_{x})^{2}/h\right\} _{x}/\text{\ensuremath{\eta_{x}}}.
\]
Then multiplying (\ref{eq:crc-sol}) with $\eta_{x}$ and using (\ref{eq:vlm-tw})
to substitute $c\eta_{x},$ after a few rearrangements, the resulting
equation can be integrated once more to obtain
\[
A\eta+B-\tfrac{1}{4}(1-\eta^{2})(1+\vartheta^{2})=-\frac{c^{2}}{12}\frac{1-2\eta_{0}\eta+\eta_{0}^{2}}{1-\eta^{2}}\text{\ensuremath{\eta_{x}}}^{2},
\]
where $B$ is a constant of integration which represents another conserved
quantity, the flux of impulse. Finally, using (\ref{eq:vlm-sol})
to eliminate $\vartheta,$ we obtain 
\begin{equation}
\text{\ensuremath{\eta_{x}}}^{2}+P(\eta)=0,\label{eq:etax}
\end{equation}
 where 
\begin{equation}
P(\eta)=-\frac{3}{c^{2}}\frac{(1-\eta^{2})(1-\eta^{2}-4(A\eta+B))+4c^{2}(\eta-\eta_{0})^{2}}{1-\eta^{2}+(\eta-\eta_{0})^{2}}\label{eq:Png}
\end{equation}
is analogous to potential energy when $\text{\ensuremath{\eta_{x}}}^{2}$
is interpreted as kinetic energy with $x$ representing the time.
Thus (\ref{eq:etax}) can be thought of as describing the conservation
of total energy for a body performing periodic oscillations in the
potential well defined by (\ref{eq:Png}). The oscillations occur
between the points at which the velocity-like quantity $\eta_{x}$
changes sign. This happens at the turning points $\eta_{x}=0$ which
correspond to two extremes of $\eta.$ According to (\ref{eq:etax}),
these points are the zeros of the numerator of $P(\eta).$ As the
latter is a quartic function of $\eta$, it can have in general four
roots. For a finite-amplitude solution to be possible, all roots have
to be real \citep{Esler2011}. If two roots coincide but the other
two are distinct, the oscillation period becomes infinite resulting
in the so-called solitary wave. If also the other two roots coincide
so that there are two distinct double roots, we end up with only a
half-solitary wave which is termed the solibore \citep{Esler2011}.

In the following, we focus on the last case and determine the unknown
constants in (\ref{eq:Png}) by factorizing the numerator of $P(\eta)$
as follows
\begin{equation}
(1-\eta^{2})(1-\eta^{2}-4(A\eta+B))+4c^{2}(\eta-\eta_{0})^{2}=(\eta-\eta_{+})^{2}(\eta-\eta_{-})^{2},\label{eq:fact}
\end{equation}
where $\eta_{+}$ and $\eta_{-}$ are two double roots of the quartic
defining respectively the downtream and upstream interface heights.
 Comparing the coefficients at the same powers of $\eta$ on both
sides of (\ref{eq:fact}) and eliminating $A$ and $B$, we have 
\begin{equation}
4c^{2}\left(1\pm\eta_{0}\right)^{2}=\left(\eta_{+}+\eta_{-}\pm(1+\eta_{+}\eta_{-})\right)^{2},\label{eq:ceta}
\end{equation}
which represents a system of two equations corresponding to the same
choice of sign on both sides. There are two solutions for $c$ and
$\eta_{0}$ possible depending on the combination of signs when taking
the square root of both sides of (\ref{eq:ceta}). The first solution
\[
c=\pm\tfrac{1}{2}(1+\eta_{+}\eta_{-}),\quad\eta_{0}=\frac{\eta_{+}+\eta_{-}}{1+\eta_{+}\eta_{-}},
\]
with $(\vartheta_{+},\vartheta_{-})=\mp(\eta_{-},\eta_{+})$ following
from (\ref{eq:vlm-sol}) corresponds to the conservative jumps with
continuous characteristic velocities found in the previous section.
The second solution 
\[
c=\pm\tfrac{1}{2}(\eta_{+}+\eta_{-}),\quad\eta_{0}=\frac{1+\eta_{+}\eta_{-}}{\eta_{+}+\eta_{-}},
\]
with $\vartheta_{+}=\vartheta_{-}=\mp1$ resulting from (\ref{eq:vlm-sol})
corresponds to the conservative jumps with discontinuous characteristic
velocities. This type of solibores seems to be of little practical
relevance because of the very specific shear values required for their
existence.

Using the factorization of the numerator of $P(\eta)$ defined by
(\ref{eq:fact}), we can integrate (\ref{eq:etax}) analytically as
follows
\begin{align}
x(\eta) & =\frac{c}{\sqrt{3}}\int\frac{q(\eta)\,\mathrm{d}\eta}{(\eta-\eta_{+})(\eta-\eta_{-})}\nonumber \\
 & =\frac{2c}{\sqrt{3}(\eta_{+}-\eta_{-})}\left[q(\eta_{+})\text{arctanh}\left(\frac{q(\eta)}{q(\eta_{+})}\right)-q(\eta_{-})\text{arctanh}\left(\frac{q(\eta_{-})}{q(\eta)}\right)\right],\label{eq:xeta}
\end{align}
where $q(\eta)=\sqrt{1-\eta^{2}+(\eta-\eta_{0})^{2}}$ and $\eta_{+}$
and $\eta_{-}$ are ordered so that $q(\eta_{+})\ge q(\eta_{-}).$
In the symmetric continuous case $\eta_{+}=-\eta_{-},$ when $\eta_{0}=0$
and $c=\pm(1-\eta_{+}^{2})/2,$ (\ref{eq:xeta}) can be written explicitly
as $\eta(x)=\eta_{+}\tanh\left(\frac{2\sqrt{3}x}{\eta_{+}-\eta_{+}^{-1}}\right).$

\begin{figure}
\begin{centering}
\includegraphics[width=0.75\columnwidth]{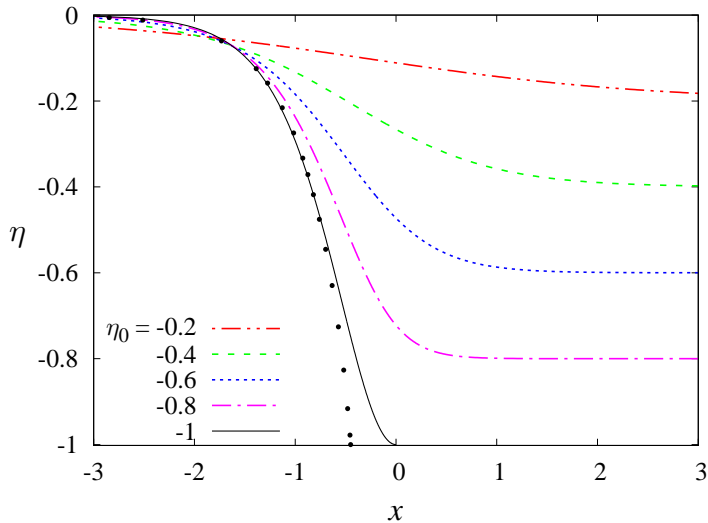}
\par\end{centering}
\caption{\label{fig:xeta}Solibores advancing into a quiescent downstream state
with various interface heights $\eta_{0}<0.$ The dots show the approximate
solution of \citet{Benjamin1968} for energy-conserving gravity current.}
\end{figure}

For a quiescent downstream state, when $\vartheta_{+}=\eta_{-}=0,$
$\eta_{0}=\eta_{+}$ and $c=\pm\frac{1}{2},$ which is practically
the most relevant, (\ref{eq:xeta}) is plotted in Fig. \ref{fig:xeta}a
for various interface heights $\eta_{0}<0.$ Note that (\ref{eq:xeta})
is invariant with the symmetry transformation $\eta\rightarrow-\eta.$
Therefore, the solutions with $\eta_{0}>0$ are mirror symmetric images
of those with $\eta_{0}<0$ shown in Fig. \ref{fig:xeta}. It is important
to note that the solution for $\eta_{0}=-1$ corresponds to the limit
of an infinitesimally shallow bottom layer. As seen, the solibore
shape in this limiting case is very similar to that of the gravity
current found by \citet{Benjamin1968}. The difference is mostly at
the bottom which is approached tangentially by the bore whereas the
gravity current approaches it at the angle of $\pi/3$ \citep{VonKarman1940}.
This exact inviscid solution, which contains a singularity at the
contact point where the velocity varies as $\sim r^{-1/2}$ with the
distance $r,$ is outside the scope of the SW approximation.

\section{\label{sec:Sum}Summary and discussion}

In the present paper, we considered a special type of hydraulic jumps
which, in the two-layer Boussinesq system bounded by a rigid lid,
can conserve not only the mass and impulse, as usual, but also the
circulation and energy. This is possible only at certain combinations
of the upstream and downstream states for which the speed of propagation
becomes invariant with the free parameter specifying the contribution
of each layer to the interfacial pressure distribution in the generalised
SW momentum conservation law.

Jumps propagating into a quiescent fluid are fully conservative only
if the interface upstream is located at the mid-height of the channel.
The velocity of propagation of such jumps is independent of their
height and equal to $\dot{\xi}=0.5$ in the usual dimensionless units.
These fully conservative jumps have a special symmetry with the upstream
shear being equal to the downstream interface height and vice versa:
$\vartheta_{\pm}=\eta_{\mp}.$ This symmetry ensures the automatic
conservation of the impulse and energy as well as the continuity of
both characteristic velocities across the jump. Namely, the positive
and negative characteristics emanating on one side of the jump are
parallel to the corresponding characteristics emanating on the other
side. This apparently allows the jump to propagate without expanding
or contracting.

Using this property we identified a broader class of fully conservative
symmetric jumps which propagate in shear flows with the velocity $\dot{\xi}=\pm\frac{1}{2}(1+\eta_{+}\eta_{-}).$
The $\pm$ sign here reflects the fact that, in the inviscid approximation,
these jumps can theoretically propagate in either direction. In real
fluids, which are viscous, physical considerations suggest that they
have to propagate in the direction that increases the shear. For example,
a jump leaving a viscous fluid right behind it at rest would be unphysical.

We also found that if the shear upstream and downstream is marginally
hyperbolic, i.e. $\vartheta_{-}=\vartheta_{+}=\mp1,$ another type
of conservative jumps are possible. These jumps are in general asymmetric,
have discontinuous characteristic velocities and propagate at the
speed $\dot{\xi}=\pm\frac{1}{2}(\eta_{+}+\eta_{-}).$ Because of the
very specific shear values required for the existence of this type
of jumps, they appear of little practical relevance.

It is important to note that the fully conservative jumps considered
in this study satisfy four conditions but constrain only three out
of five jump parameters. Namely, such jumps still have two degrees
of freedom -- the upstream and downstream interface heights. This
is obviously due to the $\eta\leftrightarrow\vartheta$ symmetry which
is an exclusive feature of the Boussinesq approximation enabling the
automatic conservation of two quantities.

Finally, the characteristics of fully conservative jumps were shown
to be the same as those of the so-called solibores which appear as
permanent-shape solutions in weakly non-hydrostatic approximation.
An exact analytical solution for solibores was obtained.
\begin{figure}
\begin{centering}
\includegraphics[width=0.75\columnwidth]{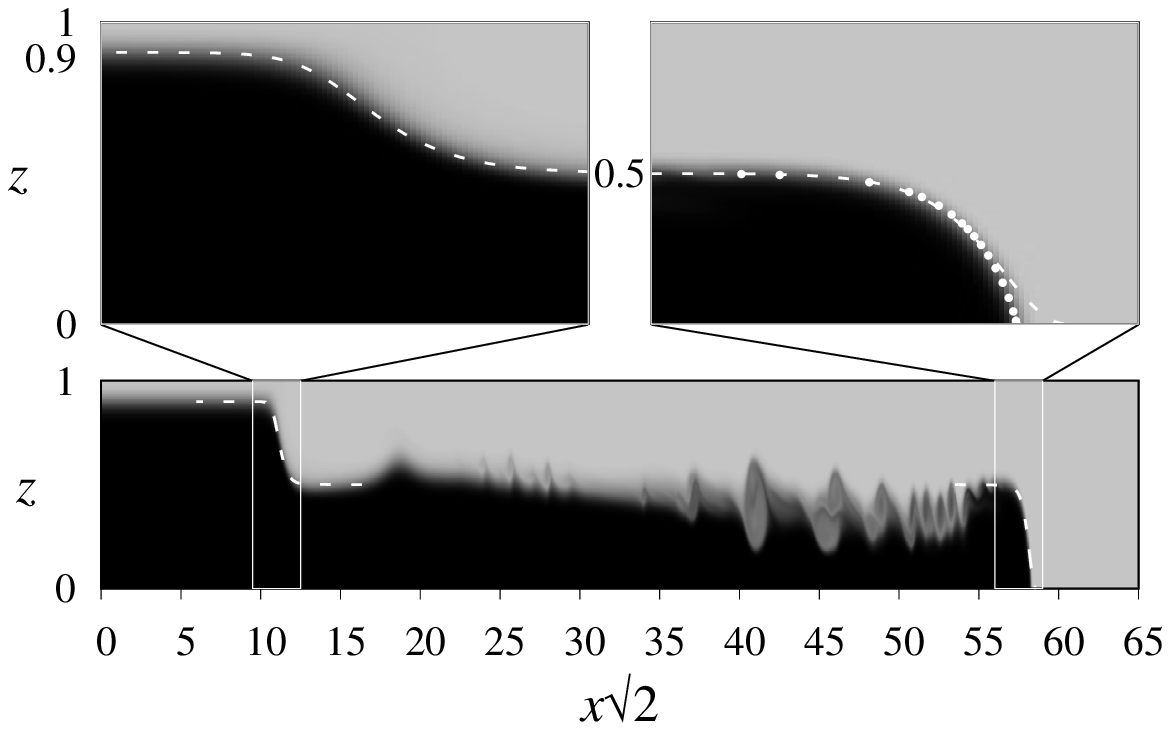}\put(-10,45){(a)}
\par\end{centering}
\begin{centering}
\includegraphics[width=0.75\columnwidth]{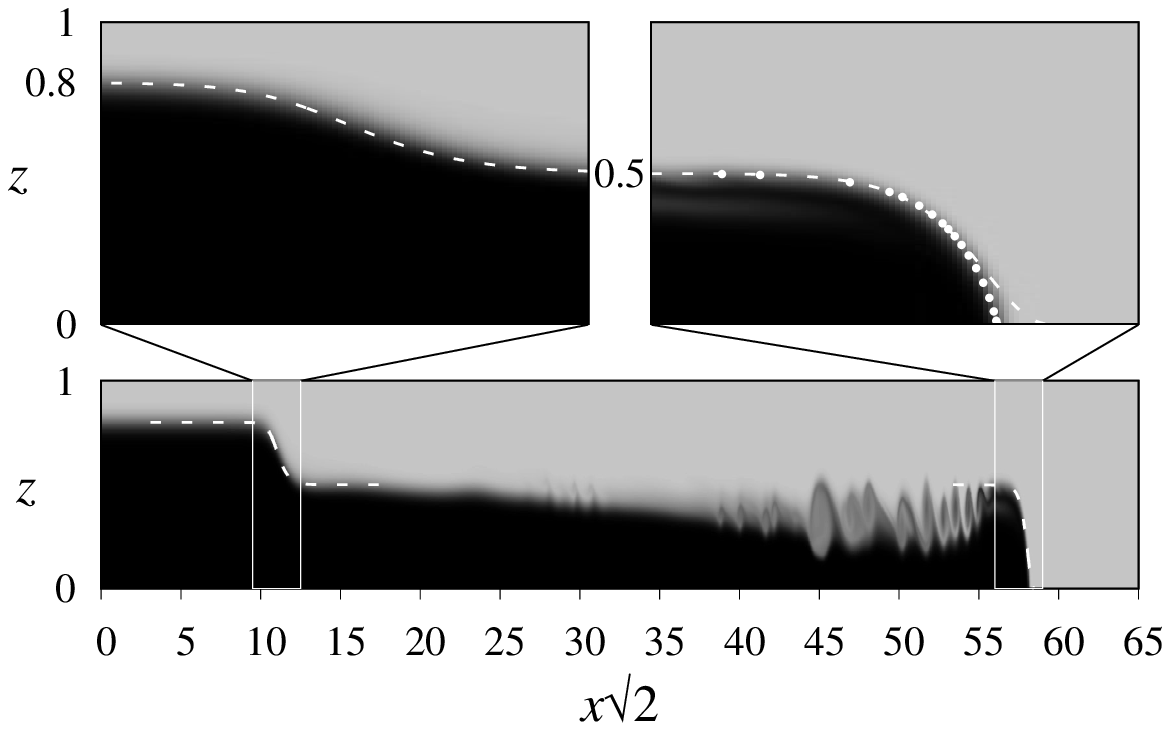}\put(-10,45){(b)}
\par\end{centering}
\caption{\label{fig:lckex}Comparison of the analytical solution (\ref{eq:xeta})
with the DNS results of \citet{Khodkar2017} for partial lock exchange
flows with the lock height $h_{-}=0.9$ (a) and $h_{-}=0.8$ (b).
The dashed lines show the analytical solution and the dots show the
approximate solution for the energy-conserving gravity current obtained
by \citet{Benjamin1968}. The DNS results shown in the background
are from figures 9(e,f) of \citet{Khodkar2017}.}
\end{figure}

Note that two such fully conservative jumps feature in the partial
lock exchange flow when the lock is taller than the channel mid-height
\citep{Politis2020}. These jumps represent the front of a gravity
current, which propagates downstream along the bottom, and a bore,
which propagates upstream in the upper half of the channel. Analytical
solution (\ref{eq:xeta}) can be seen in figure \ref{fig:lckex} to
agree well with the two-dimensional DNS results of \citet{Khodkar2017}
when the $x$-axis is scaled by a factor of $\sqrt{2}$, which is
apparently due to a different internal length scale used in the numerical
code for the $x$-coordinate. The necessity of this rescaling becomes
obvious when the numerical results of \citet{Khodkar2017} are compared
with the inviscid solution of \citet{Benjamin1968} for the energy-conserving
gravity current. This solution can be used as a benchmark because
it is known to provide a good approximation to both experimental and
numerical results \citep{Haertel2000}. Once the rescaled numerical
results match the benchmark solution for gravity current, they agree
also with the solution for solibore obtained in this study. The agreement
is very good for both heights of the lock considered. As discussed
above, the inviscid solution obtained by \citet{Benjamin1968} is
very close to the respective solibore solution except in the vicinity
of the contact point. It is interesting to note that both cases are
equivalent in the SW approximation but treated somewhat differently
in the conventional control volume approach. This, however, does not
affect the predicted speed propagation which is the same for gravity
currents \citep{Benjamin1968} and internal bores with a vanishingly
shallow downstream bottom layer \citep{Klemp1997}.

The finding that certain large-amplitude hydraulic jumps can be fully
conservative while most are not such even in the inviscid approximation,
points toward the wave dispersion as a primary mechanism behind the
lossy nature of internal bores. Namely, it is the absence of dispersion
in solibores that makes the corresponding jumps fully conservative
while all other jumps are in general dispersive \citep{Esler2011}.
Such dispersive jumps may be amenable to Whitham's modulation theory
\citep{Whitham1965,El2016} which could resolve the non-uniqueness
in their speed of propagation emerging in the hydrostatic SW approximation.

\begin{acknowledgements} Declaration of Interests. None.\end{acknowledgements}

\bibliographystyle{jfm}
\bibliography{solibo_rev-JFM}

\end{document}